\documentclass[showpacs,preprint,aps]{revtex4}%
\usepackage{amsfonts}
\usepackage{amsmath}
\usepackage{amssymb}
\usepackage{epsfig}
\usepackage{graphicx}%
\ifx\pdfoutput\relax\let\pdfoutput=\undefined\fi
\newcount\msipdfoutput
\ifx\pdfoutput\undefined\else
\ifcase\pdfoutput\else
\ifx\paperwidth\undefined\else
\ifdim\paperheight=0pt\relax\else\pdfpageheight\paperheight\fi
\ifdim\paperwidth=0pt\relax\else\pdfpagewidth\paperwidth\fi
\fi\fi\fi
\begin{document}
\title{Realizing Quantum Controlled Phase Flip through Cavity-QED}
\author{Yun-Feng Xiao}
\email{yfxiao@mail.ustc.edu.cn}
\author{Xiu-Min Lin}
\author{Jie Gao}
\author{Yong Yang}
\author{Zheng-Fu Han}
\email{zfhan@ustc.edu.cn}
\author{Guang-Can Guo}
\email{gcguo@ustc.edu.cn}
\affiliation{Key Laboratory of Quantum Information, University of Science and Technology of
China (CAS), Hefei 230026, People's Republic of China.}

\begin{abstract}
We propose a scheme to realize quantum controlled phase flip (CPF) between two
rare earth ions embedded in respective microsphere cavity via interacting with
a single-photon pulse in sequence. The numerical simulations illuminate that
the CPF gate between ions is robust and scalable with extremely high fidelity
and low error rate. Our scheme is more applicable than other schemes presented
before based on current laboratory cavity-QED technology, and it is possible
to be used as an applied unit gate in future quantum computation and quantum communication.

\end{abstract}

\maketitle

\section{Introduction}

Quantum computation based on cavity quantum electrodynamics (QED)
\cite{Yamamoto,Kimble} attracts persistent interest in experimental
realization \cite{Turchette,Chuang,Pellizzari,Cirac,Pachos}. In theory,
cavity-QED with long-lived states and high-$Q$ cavities thus provides a
promising tool for creating entanglement and superposition, and also for
implementation of quantum computing algorithms. In a number of different
schemes, quantum information usually can be represented by states of photons
\cite{Turchette,Chuang} or atomic/ionic states \cite{Pellizzari,Cirac,Pachos}.
In a classical quantum computation scheme based on cavity-QED of Ref.
\cite{Turchette}, qubits are represented by the polarized states of photons,
and high-finesse optical microcavities with atoms are used to provide
nonlinear interactions between photons. However, the storage of single-photon
information and feeding of single photons into/out of cavities are still
experimental challenges for large-scale quantum computation. In the other
case, qubits are represented by atomic states, which are ideal for the storage
of quantum information, and photons transmit information among atoms, which
are the best long-distance carriers of quantum information. An important
precondition for the case is so-called regime of strong coupling in
cavity-QED. Experimentally, the condition has been realized
\cite{Yamamoto,Kimble} or is theoretically feasible in different optical
cavities, such as micropost microcavity \cite{Fattal}, Fabry-Perot bulk
optical cavity \cite{Mckeever}, photonic crystal \cite{Painter} and
microsphere cavity \cite{Vahala}, etc. Among them, whispering gallery modes of
silica-microsphere get especial attentions because of their ultrahigh factor
$Q$ and small mode volume.

Very recently, L.-M. Duan and H. J. Kimble have proposed a new interesting
scheme to carry out quantum controlled phase flip (CPF) \cite{Duan1}, where
qubits are encoded as coherent superposition of polarized states of
single-photon pulses. They assumed $T\gg1/\kappa$, which means longer storage
time of single photons is needed, and thus it brings on a challenge for
maintaining photons coherent, and here $\kappa\ $is the decay rate of the
cavity mode field itself, and $T$ is the single-photon pulse duration. The
experimental scheme presented here tries to overcome these difficulties by
using atomic rare earth ions embedded in microsphere cavities. Qubits are
represented by hyperfine ground states of ions, which provides less storage
time of single-photon pulses and better scalability.

\section{Implementation of CPF gate}

The basic model here is first built on single three-level atoms trapped in
Fabry-Perot cavities, and single three-level ions embedded in microsphere
cavities will be discussed in the section IV later. As illustrated in Fig. 1a,
atomic states $\left\vert 0\right\rangle $\ and $\left\vert 1\right\rangle $
are two stable ground states, and state $\left\vert e\right\rangle \ $is a low
excited state. A single-photon pulse is reflected by the two cavity-atom
subsystems in sequence, and the CPF gate for the two atoms is realized by a
series of these simple reflections and some local unitary operations. The
single-photon pulse is initially prepared in an equal coherent superposition
of two orthogonal polarization components and can be expressed as $\left\vert
\phi\right\rangle _{p}=\frac{1}{\sqrt{2}}\left(  \left\vert H\right\rangle
+\left\vert V\right\rangle \right)  $. Qubits are represented by arbitrary
coherent superposition of the two atomic ground states, and the initial state
is prepared as $\left\vert \varphi\right\rangle _{12}=\left(  \beta
_{10}\left\vert 0\right\rangle _{1}+\beta_{11}\left\vert 1\right\rangle
_{1}\right)  \otimes\left(  \beta_{20}\left\vert 0\right\rangle _{2}%
+\beta_{21}\left\vert 1\right\rangle _{2}\right)  $, where $\beta_{ij}\ $is
arbitrary superposition coefficients, $i$ denotes the $atom1$ or $atom2$, $j$
denotes the state $\left\vert 0\right\rangle \ $or $\left\vert 1\right\rangle
$\ , with relations: $\left\vert \beta_{i0}\right\vert ^{2}+\left\vert
\beta_{i1}\right\vert ^{2}=1$. The atomic transition $\left\vert
1\right\rangle \rightarrow\left\vert e\right\rangle $\ is resonant with a
cavity mode of interest, which has $H$ polarization and is resonantly driven
by the $H$ polarization component of the input single-photon pulse. The CPF
gate between the atom and the photon can be described by the unitary operator
\cite{Duan1}%
\begin{equation}
U_{a,p}^{CPF}=e^{i\pi\left\vert 0\right\rangle _{i}\left\langle 0\right\vert
\otimes\left\vert H\right\rangle _{p}\left\langle H\right\vert }. \label{1}%
\end{equation}
From Eq. (1), $\left\vert 0\right\rangle $ component obtains a phase of
$e^{i\pi}$ while $\left\vert 1\right\rangle $ component keeps unchanged during
the interaction with $H$ polarization component of the input single-photon
pulse. It is very insensitive to the variation of the coupling rate $g$ even
if $g$ is not much higher than $\kappa$ \cite{Duan1}. The CPF gate between
$atom1$ and $atom2$,\textit{ }which is generated by combination of several
gates $U_{a,p}^{CPF}$ and single-bit rotation operations can be described by
the unitary operator
\begin{equation}
U_{12}^{CPF}=e^{i\pi\left\vert 0\right\rangle _{1}\left\langle 0\right\vert
\otimes\left\vert 0\right\rangle _{2}\left\langle 0\right\vert }. \label{2}%
\end{equation}
It is the most important unitary operator in our protocol which has the
following operator identity
\begin{equation}
U_{12}^{CPF}\left\vert \varphi\right\rangle _{12}\left\vert \phi\right\rangle
_{p}=U_{1p}^{CPF}R_{p}U_{2p}^{CPF}R_{p}U_{1p}^{CPF}\left\vert \varphi
\right\rangle _{12}\left\vert \phi\right\rangle _{p}. \label{3}%
\end{equation}
where $R_{p}$\ is a single-bit operating on the single-photon pulse, and the
transforming relations are $R_{p}\left\vert H\right\rangle =\frac{\sqrt{2}}%
{2}\left(  -\left\vert H\right\rangle +\left\vert V\right\rangle \right)  $
and $R_{p}\left\vert V\right\rangle =\frac{\sqrt{2}}{2}\left(  \left\vert
H\right\rangle +\left\vert V\right\rangle \right)  $. So the steps to realize
the CPF gate between $atom1$ and $atom2$ are as follows, and the overall
processes are shown in Fig. 1b. (i). $cavity1$ with $atom1$ reflects the input
single-photon pulse firstly. (ii). Make a rotation $R_{p}$ on the polarization
direction of the single-photon pulse via a half-wave plate $HWP1$. (iii).
$cavity2$ with $atom2$ reflects the single-photon pulse subsequently. (iv).
Make a rotation $R_{p}$ on the polarization direction of the single-photon
pulse via the other half-wave plate $HWP2$. (v). $cavity1$ with $atom1$
reflects the single-photon pulse again, and then the single-photon
pulse\ leaves the setup. At last, the state of the two atoms is expressed by
\begin{equation}
\left\vert \varphi\right\rangle _{12}^{\prime}=-\beta_{10}\beta_{20}\left\vert
0\right\rangle _{1}\left\vert 0\right\rangle _{2}+\beta_{10}\beta
_{21}\left\vert 0\right\rangle _{1}\left\vert 1\right\rangle _{2}+\beta
_{11}\beta_{20}\left\vert 1\right\rangle _{1}\left\vert 0\right\rangle
_{2}+\beta_{11}\beta_{21}\left\vert 1\right\rangle _{1}\left\vert
1\right\rangle _{2}, \label{4}%
\end{equation}
and meanwhile the single-photon pulse comes back to its initial state
$\left\vert \phi\right\rangle _{p}$.

\section{Theoretical Model and Analysis}

For the sake of clarity and concision, we discuss $U_{a,p}^{CPF}$\ for a
single-photon pulse and a cavity-atom subsystem, and CPF gate between atoms is
generated by simple orderly combination of $U_{a,p}^{CPF}$ and some local
unitary operations. The total Hamiltonian (single atom + single cavity mode +
free space) has the following form in the rotating frame \cite{Duan2} (in the
units of $\hbar=1$, and input single-photon pulse is $H$ polarized)%
\begin{align}
\boldsymbol{H}  &  =-i\frac{\gamma}{2}\left\vert e\right\rangle \left\langle
e\right\vert +g\left(  a_{H}\left\vert e\right\rangle \left\langle
1\right\vert +\left\vert 1\right\rangle \left\langle e\right\vert
a_{H}^{\dagger}\right)  +\int_{-\omega_{b}}^{\omega_{b}}d\omega\left[  \omega
b^{\dagger}\left(  \omega\right)  b\left(  \omega\right)  \right] \nonumber\\
&  +i\sqrt{\kappa/2\pi}\int_{-\omega_{b}}^{\omega_{b}}d\omega\left[  b\left(
\omega\right)  a_{H}^{\dagger}-a_{H}b^{\dagger}\left(  \omega\right)  \right]
, \label{5}%
\end{align}
where $\gamma$ is atomic spontaneous rate in state $\left\vert e\right\rangle
$; $a_{H}$ and $b\left(  \omega\right)  $ are respectively annihilation
operators for $H$ polarized photons in the cavity mode and in free-space modes
with the commutation relation: $\left[  b\left(  \omega\right)  ,b^{\dagger
}\left(  \omega^{\prime}\right)  \right]  =\delta\left(  \omega-\omega
^{\prime}\right)  $. Here $\omega_{b}$ is a frequency range around the
frequency of the cavity mode. In order to obtain the state of the system at
arbitrary time, two cases are considered:

1. The atom is in $\left\vert 0\right\rangle $ state at the beginning, then
the state at arbitrary time is described by%
\begin{equation}
\left\vert \Phi\left(  t\right)  \right\rangle =\left\vert 0\right\rangle
_{atom}\left\vert vac\right\rangle _{cavity}\int_{-\omega_{b}}^{\omega_{b}%
}d\omega c_{\omega}\left(  t\right)  b^{\dagger}\left(  \omega\right)
\left\vert vac\right\rangle _{freespace}+\lambda\left(  t\right)  \left\vert
0\right\rangle _{atom}\left\vert H\right\rangle _{cavity}\left\vert
vac\right\rangle _{freespace}. \label{6}%
\end{equation}
According to Schr\"{o}dinger equation $i\partial_{t}\left\vert \Phi\left(
t\right)  \right\rangle =\boldsymbol{H}\left\vert \Phi\left(  t\right)
\right\rangle $, we have%
\begin{equation}
\left\{
\begin{array}
[c]{l}%
dc_{\omega}\left(  t\right)  /dt=-i\omega c_{\omega}\left(  t\right)
-\sqrt{\kappa/2\pi}\lambda\left(  t\right)  ,\\
d\lambda\left(  t\right)  /dt=\int_{-\omega_{b}}^{\omega_{b}}d\omega
\sqrt{\kappa/2\pi}c_{\omega}\left(  t\right)  .
\end{array}
\right.  \label{7}%
\end{equation}
Then we discretize the continuum field $b\left(  \omega\right)  $ for the
numerical simulation with the single-photon pulse state replaced by
$\left\vert \phi^{\prime}\right\rangle _{p}=\sum_{k=1}^{N}c_{k}\left(
t\right)  b_{k}^{\dagger}\left\vert vac\right\rangle $, and finally we get the
following set of equations for the coefficients%
\begin{equation}
\left\{
\begin{array}
[c]{l}%
dc_{k}\left(  t\right)  /dt=-i\omega_{k}c_{k}\left(  t\right)  -\sqrt
{\kappa\Delta\omega/2\pi}\lambda\left(  t\right)  ,\\
d\lambda\left(  t\right)  /dt=\sqrt{\kappa\Delta\omega/2\pi}%
{\textstyle\sum\limits_{k=1}^{N}}
c_{k}\left(  t\right)  ,
\end{array}
\right.  \label{8}%
\end{equation}
where $N=2\omega_{b}/\Delta\omega$, $\omega_{k}=\left[  k-\left(  N+1\right)
/2\right]  \Delta\omega$, and at time $t=0$, $\ $we have $\lambda\left(
0\right)  =0$, $c_{k}\left(  0\right)  =\sqrt{\Delta\omega}c_{\omega}\left(
0\right)  $. Shape of the input single-photon pulse is described by a Gauss
function $f\left(  t\right)  =\alpha\exp\left[  -24\left(  t-T/2\right)
^{2}/T^{2}\right]  $ ($t\in\left[  0,T\right]  $).

2. The atom is in state $\left\vert 1\right\rangle $ initially. Similar
treatments can be done, and thus we have%
\begin{equation}
\left\vert \Phi^{\prime}\left(  t\right)  \right\rangle =\left\vert
1\right\rangle \left\vert vac\right\rangle \int_{-\omega_{b}}^{\omega_{b}%
}d\omega c_{\omega}^{\prime}\left(  t\right)  b^{\dagger}\left(
\omega\right)  \left\vert vac\right\rangle +\lambda^{\prime}\left(  t\right)
\left\vert 1\right\rangle \left\vert H\right\rangle \left\vert
vac\right\rangle +\mu\left(  t\right)  \left\vert e\right\rangle \left\vert
vac\right\rangle \left\vert vac\right\rangle , \label{9}%
\end{equation}
\allowbreak and%
\begin{equation}
\left\{
\begin{array}
[c]{l}%
dc_{k}^{\prime}\left(  t\right)  /dt=-i\omega_{k}c_{k}^{\prime}\left(
t\right)  -\sqrt{\kappa\Delta\omega/2\pi}\lambda^{\prime}\left(  t\right)  ,\\
d\lambda^{\prime}\left(  t\right)  /dt=\sqrt{\kappa\Delta\omega/2\pi}%
{\textstyle\sum\limits_{k=1}^{N}}
c_{k}^{\prime}\left(  t\right)  -ig\mu\left(  t\right)  ,\\
d\mu\left(  t\right)  /dt=-ig\lambda^{\prime}\left(  t\right)  -\left(
\gamma/2\right)  \mu\left(  t\right)  ,
\end{array}
\right.  \label{10}%
\end{equation}
where $\lambda^{\prime}\left(  0\right)  =\mu\left(  0\right)  =0$,
$c_{k}^{\prime}\left(  0\right)  =c_{k}\left(  0\right)  $.

In fact, the initial atomic state in our scheme is an arbitrary coherent
superposition of the two ground states, $\left\vert \varphi\right\rangle
_{a}=\left(  \beta_{0}\left\vert 0\right\rangle +\beta_{1}\left\vert
1\right\rangle \right)  $ and the input single-photon pulse is $\left\vert
\phi\right\rangle _{p}=\frac{1}{\sqrt{2}}\left(  \left\vert H\right\rangle
+\left\vert V\right\rangle \right)  $, so final state of total system can be
expressed as
\begin{equation}
\left\vert \Xi\left(  t\right)  \right\rangle _{total}=\frac{1}{\sqrt{2}%
}\left(  \beta_{0}\left\vert \Phi\left(  t\right)  \right\rangle +\beta
_{1}\left\vert \Phi^{\prime}\left(  t\right)  \right\rangle +\beta
_{0}\left\vert 0\right\rangle \left\vert vac\right\rangle \left\vert
V\right\rangle +\beta_{1}\left\vert 1\right\rangle \left\vert vac\right\rangle
\left\vert V\right\rangle \right)  . \label{11}%
\end{equation}
Gate fidelity between the atom and the photon is%
\begin{equation}
F=\left\langle \Xi_{a,p}^{Ideal}\left(  T\right)  \right\vert \rho
_{a,p}\left(  T\right)  \left\vert \Xi_{a,p}^{Ideal}\left(  T\right)
\right\rangle , \label{12}%
\end{equation}
where $\rho_{a,p}\left(  T\right)  =Tr_{cav}\left(  \left\vert \Xi\left(
t\right)  \right\rangle \left\langle \Xi\left(  t\right)  \right\vert \right)
$ is the reduced density operator of the atom and photon, and $\left\vert
\Xi_{a,p}^{Ideal}\left(  T\right)  \right\rangle $ is the final state of atom
and photon by ideal $U_{a,p}^{CPF}$ gate, taking the following form \
\begin{align}
\left\vert \Xi_{a,p}^{Ideal}\left(  T\right)  \right\rangle  &  =\frac
{1}{\sqrt{2}}\left(  -\beta_{0}\left\vert 0\right\rangle
{\textstyle\sum\limits_{k=1}^{N}}
e^{-i\omega_{k}T}c_{k}\left(  0\right)  b_{k}^{\dagger}\left\vert
vac\right\rangle +\beta_{1}\left\vert 1\right\rangle
{\textstyle\sum\limits_{k=1}^{N}}
e^{-i\omega_{k}T}c_{k}\left(  0\right)  b_{k}^{\dagger}\left\vert
vac\right\rangle \right) \nonumber\\
&  +\frac{1}{\sqrt{2}}\left(  \beta_{0}\left\vert 0\right\rangle \left\vert
V\right\rangle +\beta_{1}\left\vert 1\right\rangle \left\vert V\right\rangle
\right)  . \label{13}%
\end{align}
It corresponds with $\lambda\left(  T\right)  =\lambda^{\prime}\left(
T\right)  =\mu\left(  T\right)  =0$, and the factor $e^{-i\omega_{k}T}$
describes the phase change due to the propagation in vacuum during time $T$.
The fidelity can be written finally
\begin{align}
F  &  =\frac{1}{2}\left\vert \xi_{1}x+\xi_{2}\left(  1-x\right)  +1\right\vert
^{2}\nonumber\\
&  =\frac{1}{4}\left(  s_{2}x^{2}+s_{1}x+s_{0}\right)  , \label{14}%
\end{align}
where $\xi_{1}=-\sum_{k=1}^{N}\left[  e^{-i\omega_{k}T}c_{k}\left(  0\right)
\right]  ^{\ast}c_{k}\left(  T\right)  $, $\xi_{2}=\sum_{k=1}^{N}\left[
e^{-i\omega_{k}T}c_{k}^{^{\prime}}\left(  0\right)  \right]  ^{\ast}%
c_{k}^{^{\prime}}\left(  T\right)  $, $x=\left\vert \beta_{0}\right\vert
^{2}=1-\left\vert \beta_{1}\right\vert ^{2}$, $s_{2}=\left\vert \xi_{1}%
-\xi_{2}\right\vert ^{2}$, $s_{1}=2\operatorname{Re}\left[  \left(  \xi
_{2}^{\ast}+1\right)  \left(  \xi_{1}-\xi_{2}\right)  \right]  $,
$s_{0}=\left\vert \xi_{2}+1\right\vert ^{2}$. The minimum of the fidelity for
$U_{a,p}^{CPF}$\ can be expressed as%
\begin{equation}
F_{\min}=\left\{
\begin{array}
[c]{l}%
\frac{1}{4}s_{0},\left(  -s_{1}/2s_{2}<0\right)  ,\\
\frac{1}{4}\left(  s_{0}-s_{1}^{2}/4s_{2}\right)  ,\left(  0\leq-s_{1}%
/2s_{2}\leq1\right)  ,\\
\frac{1}{4}\left(  s_{0}+s_{1}+s_{2}\right)  ,\left(  -s_{1}/2s_{2}>1\right)
.
\end{array}
\right.  \label{15}%
\end{equation}

\section{Simulation and Discussion}

In order to take numerical simulation for the theoretical results of the
previous section, it is necessary to consider a practicable system and take
some practical parameter estimations. First of all, we consider
silica-microsphere cavities instead of Fabry-Perot cavities for better strong
coupling conditions and physical scalability. In the silica-microsphere
cavity, whispering-gallery modes (WGMs) are supported. In a ray-optics
picture, WGMs correspond to light traveling around the equator of a
microsphere, and characterized by mode numbers $q,l,m$ and their polarizations
($TE$ or $TM$), where $q$ is the radial and$\ l,m$ are angular mode numbers
respectively. What we are most interested in is the so-called fundamental WGM
($q=1,l=m$) which corresponds to the highest quality factor $Q$ and the
smallest volume $V_{m}$. Quality factor $Q$ of WGM can reach extremely high,
up to $10^{10}$ in experiments \cite{Gorodetsky}, and thus strong coupling
conditions are more easily obtained than that by F-P cavity \cite{Vahala}. In
the recent work, one group has just mentioned a quantum computation protocol
through microsphere-cavity-assisted interaction \cite{Wanghailin}. In order to
achieve good coupling between photons and WGMs, being different from F-P
cavity (direct coupling is obtainable through one mirror of the cavity), near
field evanescent wave couplers are required to provide efficient coupling
without disturbing the high-$Q$ character of the microsphere cavity, and fiber
tapers \cite{Caiming} or stripline pedestal anti-resonant reflecting optical
waveguides (SPARROW) \cite{Laine} are usually used for critical coupling
\cite{Caiming}. Fig. 2 shows the coupling between two fiber tapers and a
microsphere, i.e. the realization of gate $U_{i,p}^{CPF}$ between an ion and a
photon. For a single $TE$ type WGM, the $H$ polarization component of the
input pulse can couple into (out of) microsphere cavity via $Taper1$
($Taper2$), while the $V$ polarization component will pass the coupling region
and cannot couple into the microsphere cavity, and in essence, it is
equivalent to the reflection by the mirror $M$. Thus the physical setup is
simpler than that of F-P cavity because it has no $C1$, $C2$ and $PBS$ and
just need to control the working-state of switches $K1$ and $K2$ with low time
precision. Combining the advantages of fiber taper and microsphere cavity, the
system not only achieves best coupling efficiency (the efficiency is up to
$99.7\%$ when critical coupling is achieved) \cite{Caiming} but also supplies
good cascadibility. Secondly, we replace the neutral atoms by atomic trivalent
rare earth ions $RE^{3+}$ (such as $Pr^{3+}$ or $Eu^{3+}$, here we adopt
$Eu^{3+}$) for longer coherence time and lower spontaneous emission rate.
Quantum computation using rare-earth ions have also been proposed by several
groups \cite{Kouichi}. Quantum information can be stored in the ground state
structure for long periods of time (up to 82 $%
\operatorname{ms}%
$ \cite{Fraval}) and it is insensitive to the electric dipole-dipole
interaction. In our scheme, the states $\left\vert 0\right\rangle $ and
$\left\vert 1\right\rangle $ are respectively $^{7}F_{0}\left(  \pm5/2\right)
$ and $^{7}F_{0}\left(  \pm3/2\right)  $,\ and $\left\vert e\right\rangle $ is
$^{5}D_{0}\left(  \pm5/2\right)  $ \cite{Longdell}.

We assume a single ion $Eu^{3+}$ lies in the inner surface of the microsphere,
which is also the primary distributed position of the fundamental WGM
($q=1,l=m$). The radii of each microsphere here is about $10$ $%
\operatorname{\mu m}%
$, and mode volume $V_{m}$\ is about $300$ $%
\operatorname{\mu m}%
^{3}$ at $\lambda_{0}=579.879$ $%
\operatorname{nm}%
$; the $Q$ factor can be up to $5\times10^{7}$ \cite{Vahala}. The maximum
coherent coupling rate of an individual ion to the resonant WGM \cite{Kimble}
is given by $g_{0}=\left(  \frac{\mu^{2}\omega_{c}}{2\hbar\varepsilon_{0}%
V_{m}}\right)  ^{1/2}\approx1.0$ $%
\operatorname{GHz}%
$; here we let $\mu\sim er_{ion}/2\sim7.5\times10^{-19}$ $%
\operatorname{C}%
.%
\operatorname{nm}%
$. Decay rate of the mode\text{ reaches }$\kappa=\omega_{0}/2Q\sim32$\text{ }$%
\operatorname{MHz}%
$\text{, and we assume single-photon pulse duration }$T$\text{ is about }$3.0$
$%
\operatorname{\mu s}%
$ for $\kappa T\gg1$. Last key parameter about strong coupling conditions is
ionic decay rate to modes other than the cavity mode of interest. It is
reasonable to assume $\gamma=1$ $%
\operatorname{kHz}%
$\ here because lifetime of spontaneous emission in the excited state of rare
earth can reach several $%
\operatorname{ms}%
$ \cite{Longdell,Kouichi,Equall}. Sum up all the above, $g\gg\kappa\gg\gamma$
can be satisfied in our scheme; i.e., strong coupling between single ion and
microsphere cavity mode can be easily realized.

Based on above parameter estimations, we prove that we have realized the CPF
gate via numerical simulations. In Fig. 3a, we show ionic phase variation
under gate $U_{i,p}^{CPF}$ for different $\omega_{k}$, $\Delta\theta
_{\omega_{k}}^{_{^{\left\vert 0\right\rangle }}}=$ $\theta_{\omega_{k}%
}^{_{\left\vert 0\right\rangle }}\left(  cavity\right)  -\theta_{\omega_{k}%
}^{_{\left\vert 0\right\rangle }}\left(  vac\right)  $ and $\Delta
\theta_{\omega_{k}}^{_{\left\vert 1\right\rangle }}=$ $\theta_{\omega_{k}%
}^{_{^{\left\vert 1\right\rangle }}}\left(  cavity\right)  -\theta_{\omega
_{k}}^{_{\left\vert 1\right\rangle }}\left(  vac\right)  $, where
$\theta_{\omega_{k}}^{_{\left\vert 0\right\rangle }}\left(  cavity\right)  $
and $\theta_{\omega_{k}}^{_{\left\vert 1\right\rangle }}\left(  cavity\right)
$\ denote final ionic phase after $H$ polarization component of the photon is
reflected by cavity-ion subsystem when the ionic state is $\left\vert
0\right\rangle $ or $\left\vert 1\right\rangle $, and $\theta_{\omega_{k}%
}^{_{\left\vert 0\right\rangle }}\left(  vac\right)  $ or $\theta_{\omega_{k}%
}^{_{\left\vert 1\right\rangle }}\left(  vac\right)  $\ are final ionic phases
in the absence of cavity-ion subsystem during time $T$ (it means photon's free
propagation in vacuum). Thus we have $\theta_{\omega_{k}}^{_{^{\left\vert
0\right\rangle }}}\left(  cavity\right)  =\arg\left[  c_{k}\left(  T\right)
\right]  $, $\theta_{\omega_{k}}^{_{^{\left\vert 0\right\rangle }}}\left(
vac\right)  =\arg\left[  e^{-i\omega_{k}T}c_{k}\left(  0\right)  \right]  $,
$\theta_{\omega_{k}}^{_{^{\left\vert 1\right\rangle }}}\left(  cavity\right)
=\arg\left[  c_{k}^{\prime}\left(  T\right)  \right]  $, $\theta_{\omega_{k}%
}^{_{^{\left\vert 1\right\rangle }}}\left(  vac\right)  =\arg\left[
e^{-i\omega_{k}T}c_{k}^{\prime}\left(  0\right)  \right]  $.\ From the two
curves, it is obvious that $\Delta\theta_{\omega_{k}}^{_{\left\vert
0\right\rangle }}$ is very close to $\pi$, and $\Delta\theta_{\omega_{k}%
}^{_{\left\vert 1\right\rangle }}$ nearly equals $0$ in the frequency range of
single-photon pulse.

We also obtain the fidelity of the CPF gate for a single ion and a single
photon. Fig. 3b shows the gate fidelity associated with different pulse
duration $T$. $F_{\min}$ can reach extremely high even when $\kappa T\sim50$.
In our estimations, $F_{\min}$ is up to $0.99998$ for $T=3.0$ $%
\operatorname{\mu s}%
$. The other important parameter of quantum logic gates in quantum computation
is error rate. The dominant noise in our scheme is photon loss during gate
operations, which is especially aroused from ionic spontaneous emission, and
leads to uncontrolled free evolution of ionic states; per contra, ions evolve
governed by Hamiltonian $\boldsymbol{H}$ when the single photon is in the
cavity. Taking a rough estimation for this case, probability of spontaneous
emission loss ($\eta$) is about $\frac{1}{2\left(  1+2g^{2}/\kappa
\gamma\right)  }\approx10^{-8}$ for per gate $U_{i,p}^{CPF}$ even when
$\left\vert \varphi\right\rangle _{ion}=\left\vert 1\right\rangle $
\cite{Duan2}. So our scheme has the ability to accomplish quantum
fault-tolerance codes (error threshold is about $10^{-5}$) \cite{Preskill} if
we neglect all classical photon loss (for instance, coupling inefficiency
between fiber tapers and microspheres).

Now we discuss some technical details of our scheme. Rare earth ions are
characterized by partially full $4f$ orbitals and their spectroscopy is
dominated by $4f^{n}\rightarrow4f^{n}$ transitions. The electrons involved in
these transitions are inside filled $5s$ and $5p$ orbitals, which screen them
from perturbations caused by the lattice \cite{Longdell,Thiel}. Once a single
ion has been embedded in a silica-microsphere, it has nearly determinate
crystal field environment, and the fluctuation of its crystal field\ is so
small that linewidths of around $100$ $%
\operatorname{Hz}%
$ for a transition in the visible have been reported \cite{Equall}. These
linewidths are right so-called homogeneous broading $\Delta\omega_{hb}$. On
the other hand, for different ions, they have different crystal field
environments and hence different optical transition frequencies. This kind of
behavior brings large inhomogeneous broading $\Delta\omega_{ib}$, up to
several $%
\operatorname{GHz}%
$ \cite{pryde}. Large inhomogeneous broading is a significant challenge for
our scheme, since single photons should be kept resonant with ionic
transitions, but obviously, it is not a natural corollary of the
aforementioned protocol. In order to get over the effect derived from this
inhomogeneous broading, we improve our protocol to be more potential. Simply,
we could only add two acousto-optical (AO) shifters in fig. 2. One (AO shifter
1) is in front of the couplers, and the other (AO shifter 2) is mounted
rearward. In other words, the single-photon pulse will be modulated in the
frequency domain by AO shifters before and after reflection by the cavity. For
instance, the central frequency of the transition ($\left\vert 1\right\rangle
\rightarrow\left\vert e\right\rangle $) is assumed $\left(  \omega_{0}%
+\delta\omega_{i}\right)  $ for the $i$th ion. The central frequency of the
incident single-photon pulse\ is $\omega_{0}$, and the frequency increases
(decreases) $\delta\omega_{i}$ after it passes through AO shifter 1 (2). In
actual quantum computers, we can fabricate a number of microspheres in which
single ions are imbedded. Then we analyse their frequency spectrums one by
one, and range the ions by their central frequencies by ascending order. Every
cavity-ion subsystem includes two additional AO shifters, and thus the
single-photon pulse with central frequency $\omega_{0}$ can sufficiently
interact with every ion embedded in the respective microsphere. Furthermore,
WGMs in microsphere have been tuned successfully by several methods
\cite{Tune}, and thus cavity modes are always able to keep resonant with ionic
transitions. On second thoughts, however, we may even design\ an elegant
scheme to use the inhomogeneous broading as a constructive factor of quantum
logic gates, instead of fighting against its destructive characteristic. For
instance, we can approximatively consider that each ion has a different
discrete linear spectrum described by $\left(  \omega_{0}+\delta\omega
_{i}\right)  $ because homogeneous broading $\Delta\omega_{hb}$ is much
narrower than inhomogeneous broading. Therefore, even single ions to be
addressed individually in the case of not knowing the position of every ion.
It may be used for addressing, or writing and reading data in future quantum computers.

\section{Summary}

In conclusion, we have described a scheme to realize quantum computation in
current laboratorial technique. Compared with other schemes, our scheme has
the following significant advantages: (a) CPF gate between ions has very high
fidelity and low error rate. Routinely, in the worst case, $F\gtrsim0.9999$
and $\eta\lesssim10^{-8}$ for $U_{i,p}^{CPF}$ are obtainable. (b) Simpler
setup but good strong coupling conditions, and we need no measurement
\cite{xiao} and shorter time delay. Our delay time in total process is only
$T$ while the time is at least $2T$ in Ref. \cite{Duan1}. (c) Our scheme is
scalable because operation times $n_{op}=\tau_{coh}/\left(  2T\right)  $ can
be expected about ten thousands, and most remarkably, microsphere cavities
themselves tend to be scalable through fiber tapers or SPARROWs. It is
stirring that every cavity-ion subsystem possibly represents a node in future
quantum information and quantum computation while single-photon pulse mediates
their interaction through fiber taper or SPARROW technique. Quantum
computation based on our scheme is more applicable in lab with current
experimental technique for we can operate single ions into microsphere
cavities more controllably and accurately.\newline

\begin{acknowledgments}
We thank Prof. Z.-W. Zhou, for his helpful advice. We also like to thank J.-L.
Cheng for his help on numerical simulation. We especially acknowledge fruitful
discussions with M.-Y. Ye. This work was funded by National Fundamental
Research Program of China (2001CB309300), the Innovation Funds of Chinese
Academy of Sciences.
\end{acknowledgments}

\begin{figure}[ptb]
\includegraphics[scale=0.5]{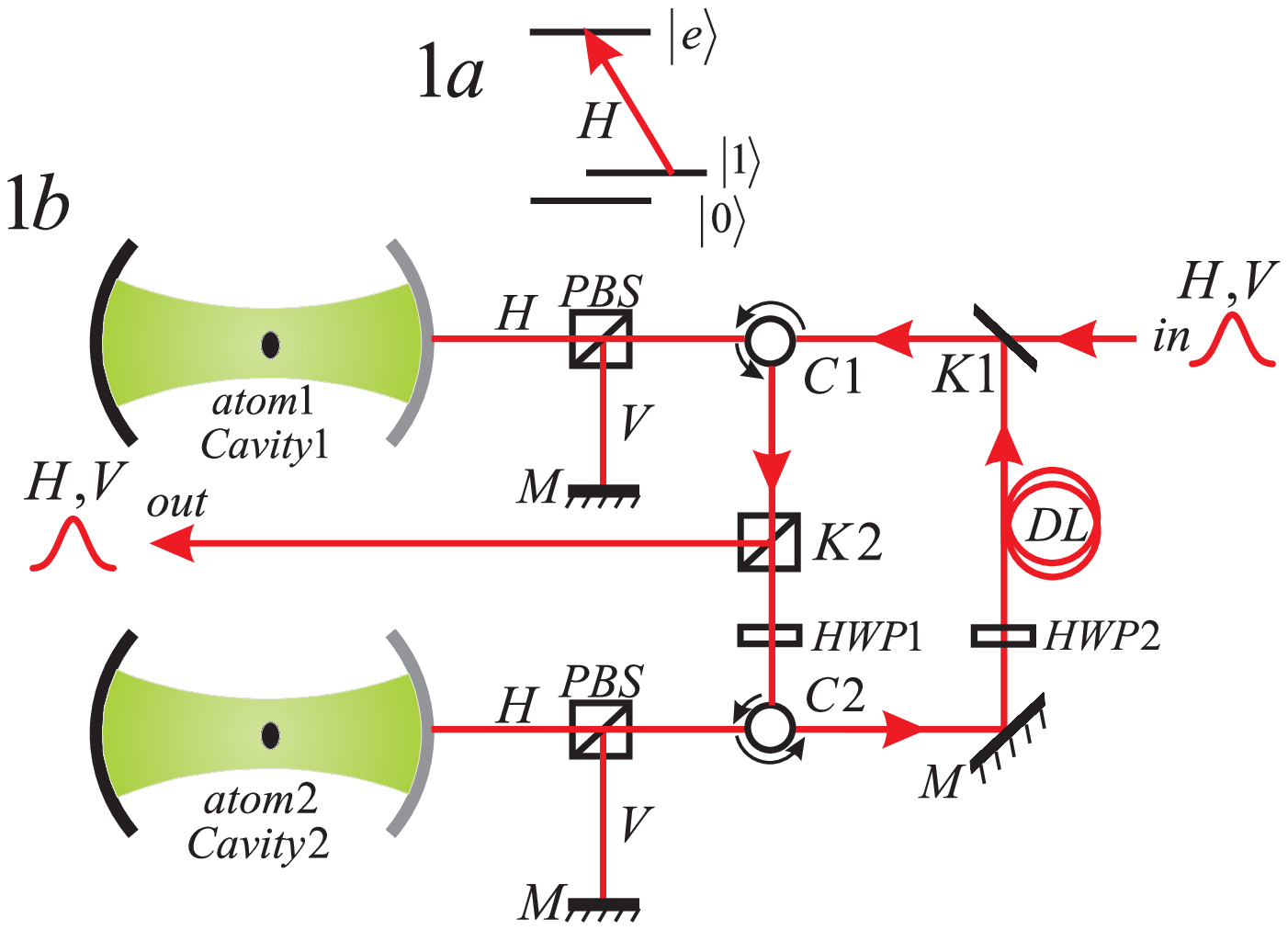}\caption{(a) The energy level diagram of
atoms. $\left\vert 1\right\rangle \rightarrow\left\vert e\right\rangle $\ is
resonant with the bare cavity mode. (b) Schematic setup to realize CPF gate
between $atom1$ and $atom2$. Both atoms lie in the two microcavities $cavity1$
and $cavity2$, respectively; $HWP1$ and $HWP2$ are two half-wave plates; $DL$
is time delay setup, for instance, fiber loops with the storage time $T$. At
time $t=0$, the working-state of switches $K1$ and $K2$ is transmitted and
kept until time $t=T$; at time $t=T$, $K1$ and $K2$ are on reflected-state;
$C1$ and $C2$ are two circulators.}%
\end{figure}

\begin{figure}[ptb]
\includegraphics[scale=0.5]{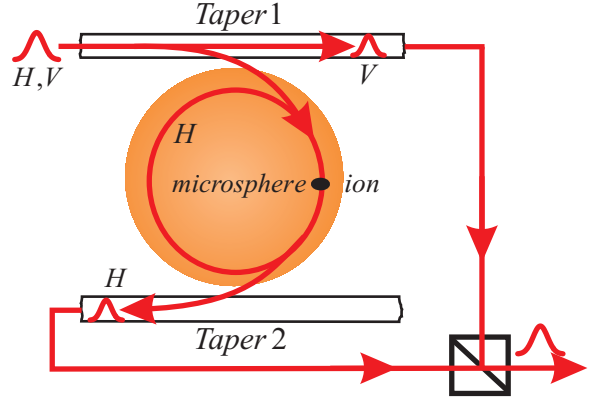}\caption{Coupling between two fiber
tapers and a microsphere, i.e. the realization of gate $U_{i,p}^{CPF}$.
$Taper1$ ($Taper2$) is input (output) coupler for $H$ polarization component.}%
\end{figure}

\begin{figure}[ptb]
\includegraphics[scale=1.0]{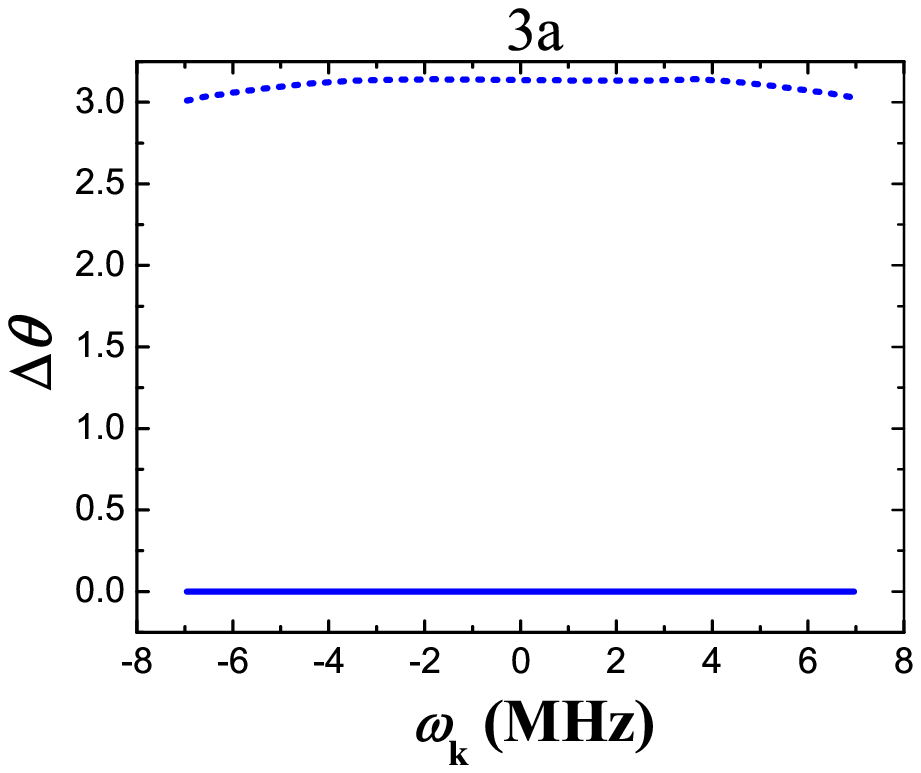}\end{figure}

\begin{figure}[ptb]
\includegraphics[scale=1.0]{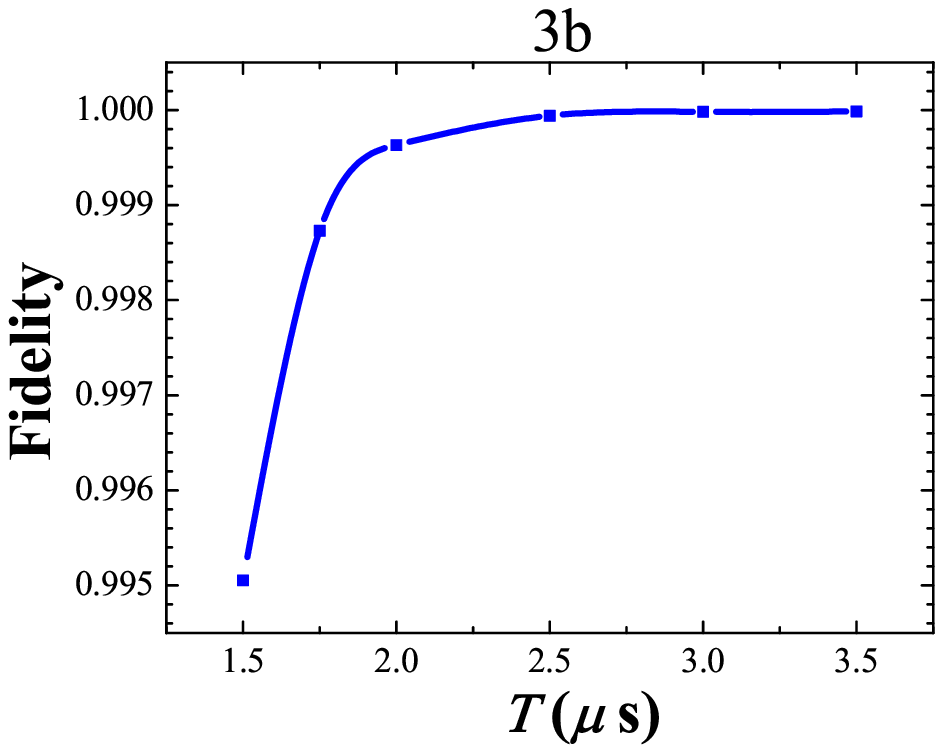}\caption{{}(a) Phase variation after the
single-photon pulse ($T=3$) is reflected. Dashed (solid) curve depicts the
phase change $\Delta\theta_{\omega_{k}}^{_{\left\vert 0\right\rangle }}$
($\Delta\theta_{\omega_{k}}^{_{\left\vert 1\right\rangle }}$) when the ion is
in $\left\vert 0\right\rangle $ ($\left\vert 1\right\rangle $). (b) Gate
fidelity between a single ion and a single photon for different pulse duration
$T$. Parameters for (a), (b), $g=1.0\operatorname{GHz}$, $\kappa
=32\operatorname{MHz}$, $\gamma=1\operatorname{kHz}$.}%
\end{figure}

\end{document}